\documentclass[conference]{IEEEtran}
\IEEEoverridecommandlockouts
\usepackage{cite}
\usepackage{amsmath,amssymb,amsfonts}
\usepackage{algorithmic}
\usepackage{textcomp}
\usepackage{xcolor}
\usepackage{url}
\usepackage{graphicx}
\usepackage[utf8]{inputenc}

\def\BibTeX{{\rm B\kern-.05em{\sc i\kern-.025em b}\kern-.08em
    T\kern-.1667em\lower.7ex\hbox{E}\kern-.125emX}}

\usepackage[whole]{bxcjkjatype} 
\usepackage[unicode]{hyperref} 

\begin{document}

\makeatletter 
\newcommand{\linebreakand}{%
  \end{@IEEEauthorhalign}
  \hfill\mbox{}\par
  \mbox{}\hfill\begin{@IEEEauthorhalign}
}

\newcommand*{\red}[1]{\textcolor{red}{#1}}
\newcommand*{\blue}[1]{\textcolor{blue}{#1}\\}

\makeatother 

\title{Quantifying Collective Emotions: Japan's Societal Trends Through Enhanced Sentiment Index Using POMS2 and SNS}

\author{\IEEEauthorblockN{Koutarou Tamura}
\IEEEauthorblockA{\textit{Center for Strategic Management \& Innovation} \\
\textit{Nomura Research Institute, Ltd.}\\
Tokyo, Japan \\
k4-tamura@nri.co.jp, k.tamura.phd@gmail.com}
\and
\IEEEauthorblockN{Yukie Sano}
\IEEEauthorblockA{\textit {Institute of Systems and Information Engineering,} \\
\textit{University of Tsukuba}\\
Ibaraki, Japan \\
sano@sk.tsukuba.ac.jp}
\and
\linebreakand
\IEEEauthorblockN{Junichi Shiozaki}
\IEEEauthorblockA{\textit{Center for Strategic Management \& Innovation} \\
\textit{Nomura Research Institute, Ltd.}\\
Tokyo, Japan \\
j1-shiozaki@nri.co.jp}
}

\onecolumn{
©2024 IEEE. Personal use of this material is permitted. Permission from IEEE must be obtained for all other uses, in any current or future media, including reprinting/republishing this material for advertising or promotional purposes, creating new collective works, for resale or redistribution to servers or lists, or reuse of any copyrighted component of this work in other works.}

\twocolumn{}

\maketitle

\begin{abstract}
In this study, we constructed an emotion index that quantitatively represents the collective emotions present in the Japanese web space by utilizing Social Networking Service (SNS) post data. Building upon previous research that used blog data and the Profile of Mood States (POMS), we restructured the methodology using posts from X (formerly Twitter) and updated the model by adding the ``Friendliness" indicator from the POMS2 metrics. Through periodic and trend analyses of the emotional indicators derived from X's post data, we found that the extension is consistent with results previously reported using blog data. This suggests that our methodology effectively captures typical emotional fluctuations in Japanese society, independent of specific SNS platforms, and is expected to serve as an index to visualize societal trends.
\end{abstract}

\begin{IEEEkeywords}
Social Network Service, Sentiment Analysis, Time Series Analysis
\end{IEEEkeywords}

\section{Introduction}
Social Networking Service (SNS) posts capture real-time reactions to societal events, and consequently, research utilizing large-scale social data has increased\cite{Lazer2009}. These data are essential for statistically analyzing the dynamics of public opinion formation and information dissemination. 

Users on SNS platforms often respond instantly to news and events. This immediate reaction is valuable for visualizing societal trends, as demonstrated by numerous studies that use SNS data to track such responses\cite{Mendoza2010,Golder2011}. For instance, research has shown how false information and its corrections spread on Twitter during the Great East Japan Earthquake\cite{Takayasu_Dema}, as well as how users on reacted to the spread of COVID-19 on Chinese Weibo\cite{Wu} and Japanese Twitter (now X)\cite{Toriumi}. However, while these studies effectively analyze reactions to specific incidents, they face challenges when comparing and evaluating responses across different events.

Some researchers have focused on developing general-purpose indices that can continuously measure social and economic dynamics, enabling cross-event analysis. Several studies have created such indices to monitor societal trends over time.
For example, Bollen et al.\cite{Bollen,Bollen2} proposed the vigor index, built using SNS data and based on the Profile of Mood States\cite{POMS_enqutte} (POMS)\footnote{The Profile of Mood States (POMS) is a psychological assessment tool developed in 1971 by Douglas M. McNair and colleagues to measure emotional states. It has been widely used in fields such as medical counseling, workplace stress management, and mental health in sports. An updated version, POMS2, has also been developed.}, which has been shown to significantly influence stock returns. Similarly, other studies have used POMS to visualize sentiment from Japanese blog data and analyze collective emotions over a 10-year period\cite{POMS_PLOS}. 
The Hedonometer\cite{HEDONO,Dodds2011} measures happiness based on words in Twitter posts and publishes results in real-time, using a binary scale of positive or negative sentiment. This method has been expanded to include languages beyond English.

In this study, we construct real-time, high-frequency indices using user post data from Twitter. Rather than relying on traditional positive-negative sentiment evaluations, we introduce more detailed emotional indicators to capture both specific events and broader social dynamics.

\section{Related Research}
We built our sentiment index based on POMS\cite{POMS_enqutte} to visualize collective emotions in society. 
POMS is a questionnaire-based method that categorizes individuals' psychological states into six emotional categories: ``Anger," ``Confusion," ``Depression," ``Fatigue," ``Tension," and ``Vigor." 
POMS can measure emotions such as ``Fatigue'' and ``Tension'' which are not covered by other indicators. 
Furthermore, its validity has already been demonstrated in prior studies conducted in both English\cite{Bollen2} and Japanese\cite{POMS_PLOS}, making it suitable for the purposes of this study.
Therefore we developed a new sentiment index for understanding and visualizing the overall emotional state of society based on POMS.

In the previous studies\cite{POMS_PLOS,POMS_origin}, emotional words were selected for each emotional category, and an emotional word dictionary was constructed. The six emotional indicators were then calculated by normalizing the ratio between the daily posting time series of each emotional word $w_{k}(t)$ from the set of emotion $e$'s emotional words $E_{e}$, and the total number of posts $w_{all}(t)$, resulting in $I_{e}(t)$:

\begin{eqnarray} r_{e}(t) &=& \sum_{k \in E_{e}}\left(\frac{w_{k}(t)}{w_{all}(t)} \right) \\
I_{e}(t) &=& \frac{r_{e}(t)- \langle r_{e}(t) \rangle }{\sigma_{e}}
\end{eqnarray}

Here, $\langle \cdot \rangle$ denotes the average, and $\sigma_{e}$ represents the standard deviation of the ratio $r_{e}(t)$. This indicator has been used successfully to visualize the collective emotions of Japanese society.

\begin{figure}[tb] \centering \includegraphics[width=\linewidth]{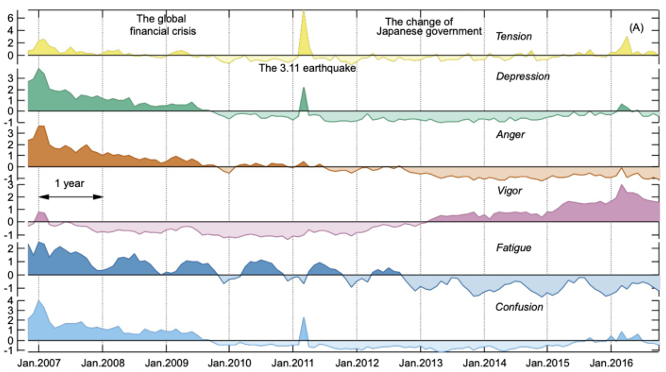} \caption{POMS-based sentiment index proposed by Sano, Fig.1(A) in \cite{POMS_PLOS}} \label{fig1} \end{figure}

It demonstrated the use of this emotional time series $I_e(t)$ during the Great East Japan Earthquake on March 11, 2011. Immediately after the earthquake, the ``Tension" indicator significantly increased as a collective emotion on blog platforms, followed by increases in the ``Depression" and ``Confusion" indicators (Fig.~\ref{fig1}). Similar emotional responses have been observed during other incidents and accidents in society.

In addition, the periodicity of the emotional indicator time series was statistically demonstrated, showing a distinct weekly cycle with significant differences between weekends and weekdays. Seasonal cycles were also observed in the ``Tension" and ``Fatigue" indicators (Fig.~\ref{fig2}).

\begin{figure}[htbp] \centering \includegraphics[width=\linewidth]{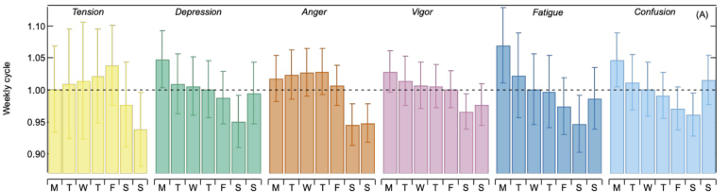} \caption{Periodical properties of the sentiment index, Fig.2(A) in \cite{POMS_PLOS}} \label{fig2} \end{figure}

Studies that apply POMS to large-scale blog data have laid the foundation for technologies that visualize societal emotional states. However, these studies face limitations due to the nature of blog platforms, such as weaker connections between users. Additionally, since POMS was originally developed for mental health diagnostics, it tends to emphasize negative emotions. An updated version, POMS2, introduces the ``Friendliness" category, providing a more nuanced classification of emotional states.

Consequently, research on societal emotional dynamics has increasingly shifted to platforms like X (formerly Twitter), which offer higher frequency, real-time data, and greater accessibility to user information. In addition, a positive-negative index is often used for simplicity.

\section{The Composition of Japan's Sentiment Index}
\subsection{Concept of Our Index}

Previous research has utilized blog post data in reference to POMS for constructing sentiment indices. However, our approach builds a sentiment index based on post data from X and integrates the emotional classifications from both POMS2 and Plutchik's Wheel of Emotions.

We reconstructed the POMS-based index using post data from X. Our sentiment index is grounded in the six POMS indicators: ``Anger," ``Confusion," ``Depression," ``Fatigue," ``Tension," and ``Vigor." Following the methodology of Sano et al.\cite{POMS_PLOS,POMS_origin}, we continue to employ the emotional word dictionary used for these six indicators.

To enhance the emotion classification, we introduce a new emotion category, ``Friendliness," from POMS2 as the seventh indicator. For this purpose, we are developing a new emotional word dictionary specifically for the ``Friendliness" indicator.

\subsection{Data Characteristics}

In this study, we use user post data from X as the primary data source. Due to the cost of data acquisition and processing, we utilize an API to count posts containing specific words, focusing on count-based acquisition rather than sentence-based extraction. From January 2016 to March 2023, data was obtained through the Twitter API, and starting from March 2023, data has been sourced from Hottolink\footnote{\url{https://www.hottolink.co.jp/} (Accessed: 2024-11-16)}. The Hottolink data consists of a 10\% sample of posts. These two data sources have been processed to ensure consistency in post count levels across them.

The collected data is limited to posts written in Japanese. As will be explained later, when extracting specific words, exclusion filters (e.g., spam-related terms) are applied to remove irrelevant posts.

\subsection{Methodology}

The pipeline for index calculation is shown in Fig.\ref{pipeline}. The key features of the method include three main processes: the construction of an emotional word dictionary, the acquisition of post count data, and the calculation of indicators.

\begin{figure*}[tb]
\centering
\includegraphics[width=\linewidth]{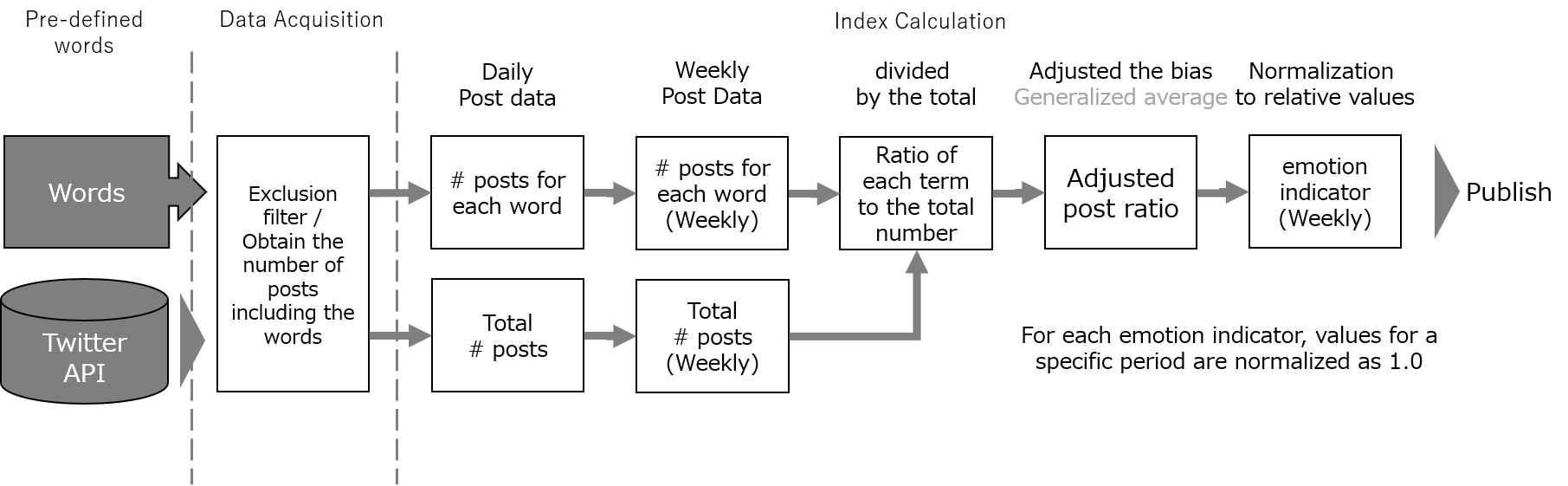}
\caption{Schematic view of the calculation flow of the sentiment index}
\label{pipeline}
\end{figure*}

\subsubsection*{Dictionary Construction}

To calculate the seven emotional indicators—``Anger," ``Confusion," ``Depression," ``Fatigue," ``Tension," ``Vigor," and ``Friendliness"—we create a ``word dictionary" by extracting emotion-related words for each indicator.

In the previous studies, this dictionary was constructed by selecting key words from the POMS questionnaire, supplemented with synonyms and spelling variations. Additionally, onomatopoeic words were included to capture a broader range of expressions. 
However, candidate words with low post counts (fewer than 5 posts per month) during 2020 were eliminated to ensure consistency.

It has shown that the time series characteristics of the indicators stabilize with approximately 25 words per emotion\cite{POMS_origin}. Therefore, we selected approximately 25 to 30 words for each emotional category.

For the six existing indicators (``Anger," ``Confusion," ``Depression," ``Fatigue," ``Tension," ``Vigor"), we adopted the word dictionary from the previous research without modification. The ``Friendliness" category, however, was newly constructed in this study, with the dictionary based on words extracted from the POMS2 questionnaire. Additionally, emotional words synonymous with influencer names (e.g., 'Waiwai（わいわい）') or names of products, services, or content were treated as outliers and excluded.

Ultimately, the number of words selected for each emotional category is as follows: 21 words for ``Vigor," 35 words for ``Confusion," 19 words for ``Depression," 25 words for ``Anger," 21 words for ``Tension," 22 words for ``Fatigue," and 24 words for ``Friendliness."

\subsubsection*{Retrieving Time Series of Post Counts}

When using the post counts of emotional words from the constructed dictionary, we exclude posts from mass media sources such as newspapers and TV stations, posts containing URLs, and replies or retweets. Additionally, we apply exclusion filters for media-related and spam-related terms across all emotional words.

Under these conditions, we obtain the time series of post counts $w_{k}(t)$ for each emotional word $k$ on day $t$ via the Twitter API. In addition to tracking individual words, we also retrieve the overall time series of post counts to capture general trends. However, since the API does not provide the total count of posts at time $t$, we retrieve the post counts for common phrases, such as ``punctuation only," which are likely to appear in any post. This serves as the total post count (hereafter referred to as the total count) $w_{all}(t)$.

\subsubsection*{Generalized Average and Normalization for Index Calculation}
To construct the time series for an emotional indicator, each emotion consists of multiple emotional words, which can introduce bias in the posting levels of certain words. 
Therefore, it is necessary to statistically adjust the data volume to prevent specific words from disproportionately influencing the indicator. 
For example, as shown in Fig.\ref{corr}, if we directly sum the post counts by their ratio in the ``Anger" indicator, certain words may dominate the results, causing the ``Anger" indicator to be overly affected by these words' trends. 
To correct this bias, we use a generalized average. Specifically, for each emotional word $k$ from the set of emotional words $E_{e}$ associated with emotion $e$, we normalize the daily post counts $w_{k}(t)$ using the overall post count time series $w_{all}(t)$, thereby reducing the influence of individual words. The emotional indicator time series for emotion $e$, $I_{e}(t)$, is then defined as follows:

First, we calculate the generalized average of the ratios:
\begin{equation}
r_{e}(t) = \left( \frac{1}{|E_{e}|}\sum_{k \in E_{e}}\left(\frac{w_{k}(t)}{w_{all}(t)} \right)^\alpha \right)^{1/\alpha}
\end{equation}
where $\alpha$ is set to $0.5$ in this study.

\begin{figure}[tb]
\centering
\includegraphics[width=\linewidth]{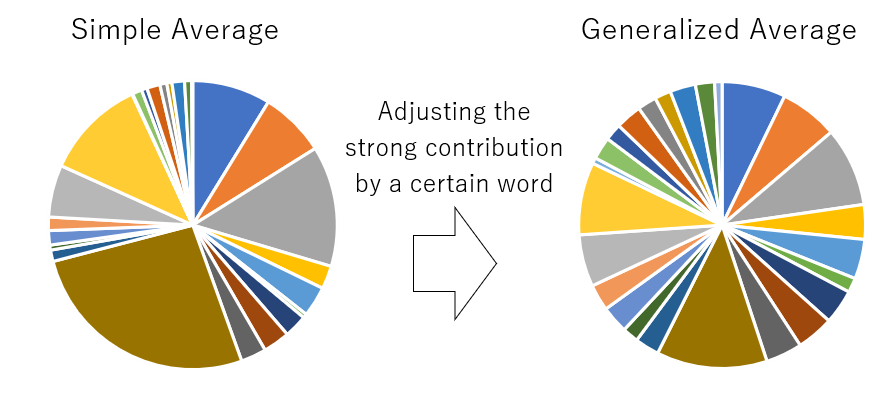}
\caption{the contribution ratio of the post counts for words defined by the ``Anger" indicator to the indicator value. The left shows the ratio of post counts to the total is averaged to construct the indicator. The right is the ratio using the generalized average with $\alpha =0.5$.}
\label{corr}
\end{figure}

Since the defined emotional indicators have different numbers of words for each emotion, the absolute values of the indicators cannot be directly compared across different emotions. To allow for meaningful comparisons of the trends across indicators, the values of each indicator are normalized by setting the average value from 2021 to 2023 as the baseline of 1.0.

\begin{equation}
I_{e}(t) = \frac{r_{e}(t)}{\langle r_{e}\rangle_{2021 \le t \le 2023}}
\end{equation}

\section{SNS Sentiment Index}
Figure \ref{events} shows each indicator calculated using our proposed method introduced in the previous section.
Specific social events that triggered significant reactions are tagged in Fig.\ref{events}, highlighting how each emotional indicator responds to various societal occurrences.

\begin{figure*}[tb]
\centering
\includegraphics[width=\linewidth]{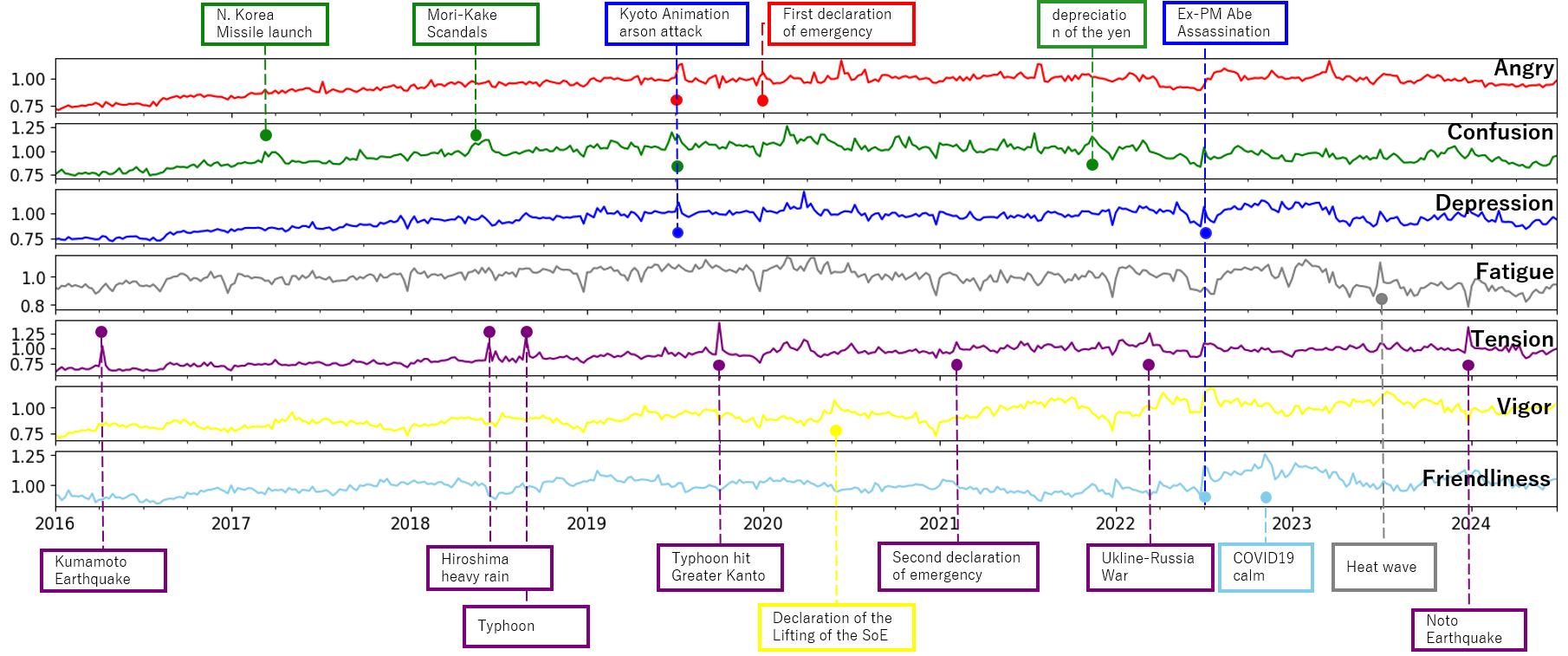}
\caption{The indicator values for each emotion in the Japan Emotion Index defined in this study. The values are normalized so that the period from 2021 to 2023 is set to 1. Dotted lines link relevant events to the points where the time series shows significant fluctuations (peaks).}
\label{events}
\end{figure*}

The ``Vigor" indicator reacts to events such as sports achievements or regulatory relaxations, while the ``Confusion" indicator responds to political events and economic news. ``Depression" fluctuates in reaction to incidents or entertainment news, and ``Tension" responds to large-scale disasters or national events. The ``Anger" and ``Fatigue" indicators are influenced by long-term policies or broader social conditions. These trends are consistent with previous research. Although the new ``Friendliness" indicator tends to be less responsive in the short term, it shows fluctuations in response to positive cultural and artistic news, such as the resolution of incidents, accidents, or the announcement of awards.

During the COVID-19 pandemic, various policies and media reports caused significant fluctuations in the indicators. For instance, during the first state of emergency, the ``Anger" indicator rose due to the delay in issuing restrictions, while the lifting of the emergency resulted in an increase in the ``Vigor" indicator. During the second state of emergency, the ``Tension" indicator increased as infections continued to spread, followed by a rise in the ``Fatigue" indicator, likely due to the extended period of self-restraint. Additionally, in October and November 2022, when the threat of COVID-19 temporarily subsided, and in May 2023, when the disease was downgraded to Category 5 in Japan, the ``Friendliness" indicator saw an increase.

As illustrated, even within the same event, such as the COVID-19 state of emergency, the emotional response can vary based on timing and context, which is evident from the indicators. The ability to break down a single event into multiple emotional dimensions provides significant insights. In particular, the newly added ``Friendliness" indicator functions as a measure of collective calmness, capturing group emotions that were not observed in previous studies. This indicator is expected to successfully capture collective emotions that were previously unmeasured.

\section{Statistical Properties of the Index}
Each emotion indicator is expected to closely follow human life patterns, and the time series is modeled to express four key features: periodicity, event effects, long-term trends, and outliers. To extract these features, we applied a simplified implementation of the Generalized Additive Model (GAM) using Prophet\cite{Prophet}.

\subsection{Periodic Components} 
We investigated the predictable periodic fluctuations and trends of the seven emotions, as well as the collective emotional responses influenced by events that could not be explained by these patterns.
The decomposition of the emotion indicators into long-term trends, periodic components, and other elements produced the following results:

\begin{figure}[tb]
\centering
\includegraphics[width=0.9\linewidth]{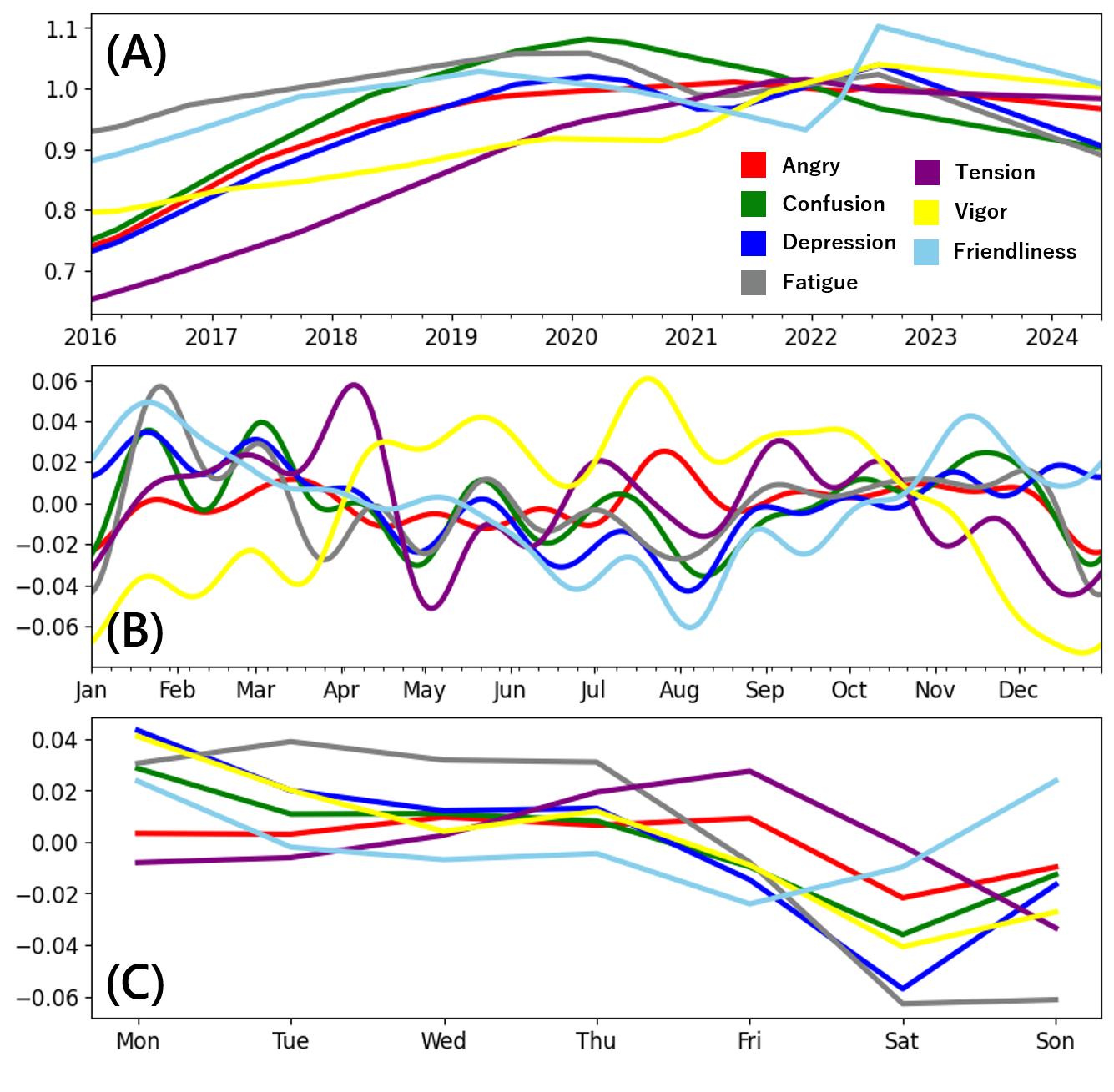}
\caption{Results of the time series analysis of the emotional indicators. (A) Long-term trends of each indicator. (B) Annual periodic fluctuations of each indicator. (C) Weekly (day-of-week) trends of each indicator.
}
\label{trend}
\end{figure}

\subsubsection*{Long-term Trend Tendencies}
In Fig.\ref{trend}(A), overall, a rising trend can be observed leading up to 2020, with the ``Vigor" and ``Tension" indicators showing particularly strong increases. During the early stages of the COVID-19 pandemic, these indicators slightly decreased, but as the pandemic subsided, both the ``Vigor" and ``Friendliness" indicators began to rise again.

\subsubsection*{Seasonal Periodic Components}
In Fig.\ref{trend}(B), most indicators display stable trends, but notable annual fluctuations are seen in the ``Tension" and ``Fatigue" indicators. The ``Tension" indicator rises at the start of the fiscal half-years, in April and September-October, coinciding with changes in living environments. The ``Fatigue" indicator tends to increase during long holidays, such as the summer break, when people's activities intensify. Conversely, the ``Depression" indicator shows minimal seasonal fluctuation, rising and falling primarily in response to specific societal events, irrespective of the time of year.

\subsubsection*{Weekly Periodic Components}
In Fig.\ref{trend}(C), all indicators are generally higher on Mondays and decrease towards the weekend, reaching their lowest levels on Saturdays and Sundays. The ``Tension" indicator shows relatively small weekly fluctuations but tends to rise toward Friday. In contrast, the ``Anger" indicator exhibits larger fluctuations, suggesting that anger, often triggered by interpersonal relationships in workplaces or schools, is more prevalent on weekdays.

It was found that the emotional indicators also fluctuate independently of specific events. Since similar trends were observed in the six indicators from prior research using blog data, there appears to be no significant differences between the blog and X post data, nor in the calculation algorithms. This consistency suggests that the indicators reliably reflect the emotional trends of SNS users. In particular, the day-to-day fluctuations are qualitatively explainable and present no inconsistencies, further supporting the potential of the proposed method as a valuable tool for visualizing societal dynamics.

\subsection{Analysis of Event-Specific Impacts}

While the indicators are designed to capture the general sentiment of Japanese society, we propose a quantitative method for analyzing the impact of individual events. Based on trend and periodicity analyses conducted using Prophet, we identified factors that fluctuate independently of specific events. By excluding these factors, we can isolate and measure the unique impact of each event.

To measure the impact of an event on a given day $T$, the time series up to $T-1$ is modeled using Prophet. For the data after $T$, the event-specific impact is extracted by subtracting the periodic and long-term trends from the actual time series. The differential ratio, $ \Delta I_{e}(t)$, represents the event-specific impact and is calculated as follows:

\begin{equation} \Delta I_{e}(t) = \frac{I_{e}(t) - \hat{I_{e}}(t)}{\hat{I_{e}}(t)} \end{equation}
where $\hat{I_{e}}(t)$ is the modeled time series using Prophet.
For example, on March 8, 2024, it was reported that the famous comic writer Toriyama Akira passed away. The raw indicator values around this date were also influenced by the annual prayers for the victims of the Great East Japan Earthquake, which led to increases in both the ``Depression" and ``Friendliness" indicators. 
However, the differential values, $ \Delta I_{e}(t)$, provide a clearer view of the specific impact of Toriyama's passing by isolating it from the overlapping event.
In a similar manner, the top-ranked specific impacts in 2024 are listed in Tab.\ref{impact}:

\begin{figure}[tb] \centering \includegraphics[width=\linewidth]{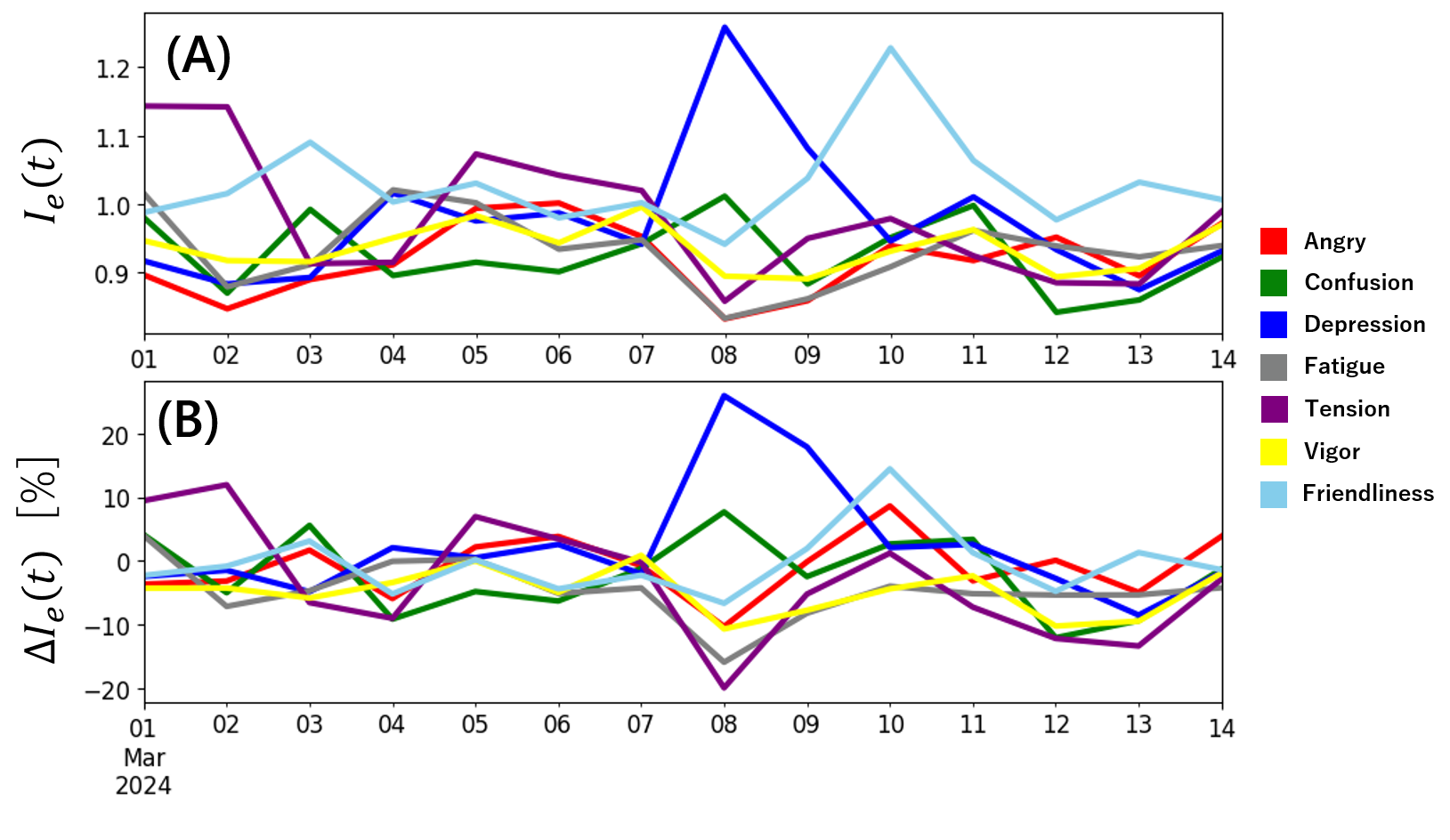} \caption{Time series of emotion indicators for March 2024. (A) Raw indicator time series. (B) Indicator time series with trends and periodic components removed.} \label{diff} \end{figure}

\begin{table}[tb]
\caption{impactful social events in 2024}
\begin{tabular}{|c|c|c|c|c|} \hline
&Emotion& Data & $\Delta I_{e}(t)$ & Detail \\ \hline \hline
1&Tension & 2024/1/1  & +58.8\% & Noto Peninsula Earthquake  \\ \hline
2&Tension & 2024/1/2  & +51.7\% & Noto Peninsula Earthquake  \\ \hline
3&Depression & 2024/3/8 & +26.0\% &  Death of Akira Toriyama \\ \hline
4&Depression  & 2024/1/2 & +22.2\% & JAL Collision Incident  \\ \hline
5&Depression & 2024/1/29 & +21.8\% & famous writer's suicide  \\ \hline
6&Tension & 2024/1/3 & +20.1\% & Noto Peninsula Earthquake  \\ \hline
7&Tension  & 2024/4/17 & +19.9\% & Bungo Strait Earthquake  \\ \hline
8&Tension & 2024/4/3 & +19.4\% & Taiwan Earthquake  \\ \hline
9&Tension & 2024/6/3 & +19.4\% &  Noto Peninsula Earthquake \\ \hline
10&Tension & 2024/3/21 & +19.2\% & Sports Betting Scandals \\ \hline
\end{tabular}
\label{impact}
\end{table}

Most of the top impactful events in 2024 were earthquakes, especially the Noto Peninsula earthquake. However, reviewing the events from an impact perspective shows that accidents and deaths, such as the passing of Akira Toriyama and the JAL collision, also shocked the Japanese public.

By excluding periodic components, this method allows for a precise quantification of event-specific impacts, enabling comparisons between events occurring at different times.

\section{Results}

In this study, we reconstructed a sentiment index using X post data, based on the prior research that used POMS. The indicators followed the methodology of previous studies, with the addition of a new ``Friendliness" indicator. An emotion index composed of seven indicators was created, and it was confirmed that these indicators responded sensitively to societal events. It was observed that even similar events could elicit different responses depending on the timing and social context, allowing for detailed emotional analysis of events.

Additionally, time-series analysis was conducted using Prophet, where the sentiment time series was decomposed into trend and periodic components. It was found that these fluctuations were similar to those reported in previous studies. The consistency of these characteristics, even with different data sources, suggests that habitual emotional fluctuations of the Japanese population are being captured. Furthermore, we proposed a method for quantifying event-specific impacts by excluding the trends and periodic components that fluctuate independently of events, enabling cross-event comparisons of different occurrences.

\section{Conclusion and Future Works}

In this study, we reconstructed a sentiment index based on prior research, successfully integrating the ``Friendliness" indicator and transitioning from POMS to POMS2. Additionally, we shifted the data source from blogs to X (formerly Twitter). We confirmed that the methodology is robust and that the index can be effectively extended to POMS2.

The indicators responded to various societal events, including notable reactions to disasters such as earthquakes, aligning with trends observed in previous research. Moreover, the newly introduced ``Friendliness" indicator, absent in earlier studies, showed expected variations, such as an increase toward the end of the COVID-19 pandemic. This indicates that the ``Friendliness'' indicator was successfully incorporated as an additional emotional dimension.

Future challenges include reassessing the comprehensiveness of the emotional categories and developing methods that combine anomaly detection techniques to automate the construction of emotional word dictionaries. Furthermore, linking specific events to the time-series analysis of the index through external information will be essential for a deeper understanding of the index's behavior.

Part of this methodology has already been adopted as the ``NRI Sentiment Index," proposed by Nomura Research Institute, and it is expected that this index will be widely used to visualize societal trends.


\end{document}